\documentclass[aps,twocolumn,nofootinbib]{revtex4}
\usepackage{amsmath,epsfig}

\newcommand{\dv}{d_V}
\newcommand{\da}{d_A}
\renewcommand{\mtt}{m_{t \bar t}}
\newcommand{\mjj}{m_{jj}}
\newcommand{\amc}{{\sc Mad\-Graph5\textunderscore}a{\sc MC@NLO}}
\def\be{\begin{equation}}
\def\ee{\end{equation}}
\def\bsp#1\esp{\begin{split}#1\end{split}}

\begin{document}

\leftline{}
\rightline{CERN-PH-TH-2014-259}
\title{Pinning down top dipole moments with ultra-boosted tops}

\author{
  Juan A. Aguilar--Saavedra$^{(a)}$,
  Benjamin Fuks$^{(b,c)}$
  and Michelangelo L. Mangano$^{(c)}$
}

\affiliation{{\phantom.}\\
$^{(a)}$ \mbox{Departamento de F\'{\i}sica Te\'orica y del Cosmos, Universidad
  de Granada, E-18071 Granada, Spain}\\
$^{(b)}$ \mbox{Institut Pluridisciplinaire Hubert Curien/D\'epartement Recherches Subatomiques}, 
    Universit\'e de Strasbourg/CNRS-IN2P3, 23 Rue du Loess, F-67037 Strasbourg, France\\
$^{(c)}$ \mbox{CERN, PH-TH, CH-1211 Geneva 23, Switzerland}
}

\begin{abstract}
We investigate existing and future hadron-collider constraints on the
top dipole chromomagnetic and chromoelectric moments, two quantities
that are expected to be modified in the presence of new physics. We
focus first on recent measurements of the inclusive top pair
production cross section at the Tevatron and at the Large Hadron
Collider. We then analyse the role of top-antitop events produced at
very large invariant masses, in the context of the forthcoming
13-14~TeV runs of the LHC, and at a future 100 TeV proton-proton
collider. In this latter case, the selection of semileptonic decays to hard muons allows
to tag top quarks boosted to the multi-TeV regime, strongly reducing
the QCD backgrounds and leading to a significant improvement in the
sensitivity to anomalous top couplings. \end{abstract}


\maketitle

\section{Introduction}
The first run of the Large Hadron Collider (LHC) at CERN has successfully
confirmed the Standard Model (SM) of particle physics with the discovery of a
Higgs boson with SM-like properties~\cite{Aad:2012tfa,Chatrchyan:2012ufa}.
Although no sign of physics beyond the SM has yet been 
observed, new
phenomena at energies not far from the electroweak scale
are predicted in many theories of new physics. There is therefore a
great expectation for the upcoming LHC runs at
center-of-mass (CM) energies \mbox{$\sqrt{s}=13$~TeV} and 14~TeV, and for
experiments at future accelerator facilities running at larger CM
energies and luminosities. Furthermore, even if
new physics were not directly reachable at the
LHC or future machines, it might still be indirectly probed through precision
measurements of the properties of the SM particles. In this context, the
top quark is  believed to play an important role,
due to the closeness of its mass to the electroweak scale. 
This has motivated an intense
research program dedicated to the study of its properties at
the LHC. Future colliders will moreover
be able to exploit the increase in CM energy and luminosity to probe indirect effects of new physics at higher momentum transfers, increasing the sensitivity to heavy new physics.

The conceptual design studies of new accelerator complexes for future
circular colliders (FCC) have recently started, at CERN~\cite{fcchh}
and at IHEP~\cite{ihep}. Their ultimate goal is the operation of a proton-proton
($pp$) collider designed to operate at
$\sqrt{s}=100$~TeV. This accelerator will allow the exploration of
energy scales several times higher than at the LHC, and will also
significantly increase the statistics of known particles.  For
example, one trillion top quarks should be available with an
integrated luminosity of 10~ab$^{-1}$. This consequently opens the
door to indirect searches of new physics in the top sector with an
unprecedented sensitivity.

In this paper we explore some of the opportunities offered by such a
huge statistics of top quarks, focusing on the sensitivity to
anomalous couplings to the gluons. 
The leading indirect effects from new physics present at a heavy scale
$\Lambda$ can be parametrized by adding
to the SM Lagrangian ${\cal L}_\text{SM}$ a set of dimension-six
operators $O_x$ invariant under the SM gauge
symmetry~\cite{Burges:1983zg,Leung:1984ni,Buchmuller:1985jz},
\begin{equation}
  {\cal L} = {\cal L}_\text{SM} + \mathcal{L}_\text{eff} = 
    {\cal L}_\text{SM} + \sum_x \frac{C_x}{\Lambda^2} O_x + \dots \ ,
\end{equation}
where the Wilson coefficients $C_x$ depend on the type of new physics and how it
couples to the SM particles. After the spontaneous breaking of the electroweak
symmetry, these operators generate corrections to the SM
couplings included in ${\cal L}_\text{SM}$, as well as interactions not present
at the tree level, such as electric dipole moments and explicit 
magnetic dipole moments. The effects on the top dipole moments can be parametrized by adding an effective
term to the top-gluon gauge coupling,
\begin{equation}
  \mathcal{L}_\text{tg} \!=\!
    - g_s \bar t \gamma^\mu \frac{\lambda_a}{2} t \, G_\mu^a
    + \frac{g_s}{m_t} \bar t \sigma^{\mu \nu} (d_V + i d_A \gamma_5)
         \frac{\lambda_a}{2} t\,  G_{\mu \nu}^a \, ,
\label{ec:lagr}
\end{equation}
with $G_{\mu \nu}^a$ ($a=1,\dots,8$) being the
gluon field strength tensor, $g_s$ the strong coupling constant, $m_t$ the top
mass and  $\lambda_a$ the Gell-Mann matrices. The second
term above contains both $gt\bar t$ and $ggt\bar t$ interactions
that arise, in the conventions of Refs.~\cite{Buchmuller:1985jz,AguilarSaavedra:2008zc}, from the
dimension-six operator
\begin{equation}
 O_{uG\phi}^{33} = (\bar q_{L3} \lambda_a \sigma^{\mu \nu} t_R) \tilde \phi
  \, G_{\mu \nu}^a\ ,
\end{equation}
where $q_{L3}$ denotes the weak doublet of left-handed quark
fields of third generation, $t_R$ the right-handed top quark field and
$\phi$ is a
weak doublet of Higgs fields (we define here
$\tilde \phi = i \tau_2 \phi^*$). After electroweak symmetry breaking, this
operator yields the top dipole moments
\begin{equation}
  d_V = \frac{\sqrt{2} v m_t}{g_s \Lambda^2} \text{Re} \, C_{uG\phi}^{33}  \,,\quad
  d_A = \frac{\sqrt{2} v m_t}{g_s \Lambda^2} \text{Im} \,C_{uG\phi}^{33} \,,
\label{ec:op}
\end{equation}
where $v=246$~GeV is the vacuum expectation value of the neutral component of
$\phi$. For weakly-interacting new physics at the TeV scale, $C_{uG\phi}^{33}\sim {\cal O}(1)$ so that one expects $d_{V,A}\sim 0.05$. This exceeds both
the chromomagnetic dipole moment generated in the SM at the one-loop
level \mbox{$\dv = -0.007$}~\cite{Martinez:2007qf} and the associated negligible
chromoelectric moment~\cite{Soni:1992tn}. 

Direct limits on the top chromomagnetic and chromoelectric dipole
moments can be derived from measurements of the $t \bar t$
inclusive cross section at the Tevatron and the LHC~\cite{Haberl:1995ek,%
Hioki:2009hm,Hioki:2013hva}, and of several $t \bar t$
differential distributions~\cite{Cheung:1995nt,HIOKI:2011xx,Kamenik:2011dk}.
Moreover, weaker bounds have been recently calculated~\cite{Bernreuther:2013aga}
from a CMS measurement of the $t \bar t$ spin correlation in LHC data at
8~TeV~\cite{CMS:2014bea}, as this observable is also modified
by the anomalous interactions of Eq.~\eqref{ec:lagr}. The most stringent limits
on $\dv$ and $\da$ however arise nowadays from low-$Q^2$ probes, such as measurements of
the neutron electric dipole moment, which constrain
\mbox{$|d_A| \leq 9.5\times 10^{-4}$}
at the 90\% confidence level (CL)~\cite{Kamenik:2011dk}, and rare $B$-meson
decays, which imply $-3.8 \times 10^{-3} \leq \dv \leq 1.2 \times 10^{-3}$ at the
95\%~CL~\cite{Martinez:2001qs}.

At the LHC running at 14 TeV, and even more at 100~TeV, a
significant amount of top-antitop pairs with a multi-TeV invariant
mass will be produced, with contributions dominated by the gluon-gluon
fusion channel. These kinematical configurations with very large
momentum transfer allow to explore the structure of the $ttg$
couplings at the shortest distances, and should then be particularly
sensitive probes of the top dipole moments~\cite{Englert:2012by,Englert:2014oea}. After revewing the
constraints that can be obtained from the measurements of total
production cross sections, in this paper we therefore focus on the
study of very high mass top-antitop final states, where the top quarks
are necessarily highly boosted. We consider a simple-minded approach
to extract the top-antitop signal from the large QCD background, and
verify that, at 100~TeV, this is sufficient to significantly push the
sensitivity to both chromoelectric and chromomagnetic dipole moments.

\section{Tevatron and LHC limits}
\begin{figure}[t]
  \begin{center}
  \includegraphics[width=.80\columnwidth,clip=]{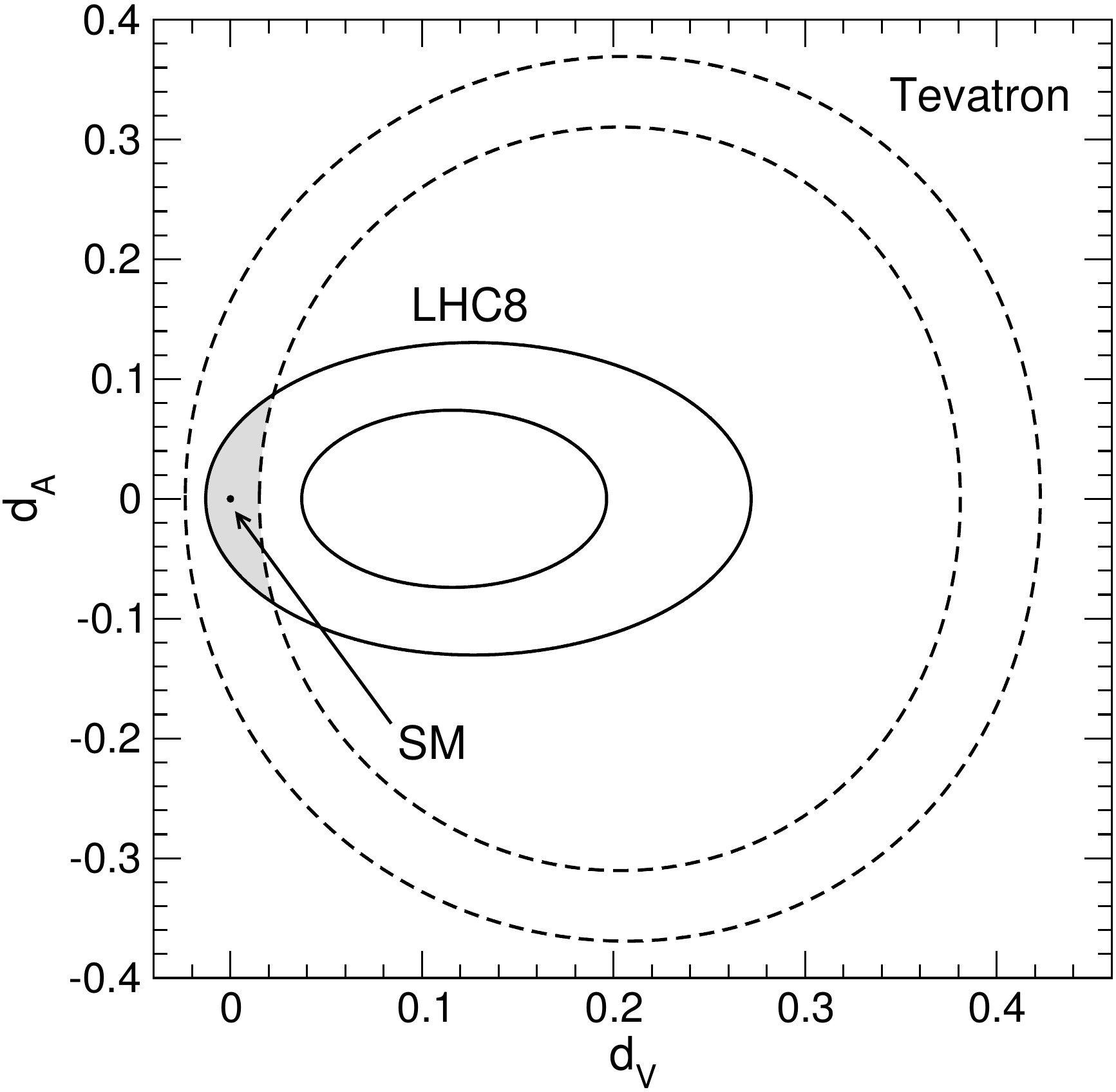}
  \caption{\small Limits, at the 95\% CL, on $\dv$ and $\da$ as extracted from
    measurements of the top-pair production total cross section at the Tevatron
    (dashed) and the LHC with \mbox{$\sqrt{s}=8$~TeV} (solid).}
\label{fig:tevlhc8}
  \end{center}
\end{figure}

The combination of inclusive $t\bar t$ cross section measurements at the Tevatron and the LHC provides much stronger limits on the top dipole moments than the individual measurements. The complementarity of these two colliders is due to a very different functional dependence of the total cross section on $\dv$ and $\da$ at the Tevatron ($p \bar p$ collisions at 1.96~TeV) and the LHC ($pp$ collisions at
7, 8~TeV), owing to the dominance of $q\bar q \to t \bar t$ at the former collider and $gg \to t \bar t$ at the latter.
Making use of the {\sc FeynRules} package~\cite{Alloul:2013bka,Degrande:2011ua}
to import the Lagrangian of Eq.~\eqref{ec:lagr} into \amc~\cite{Alwall:2014hca},
we evaluate the $t\bar t$ total production cross section at the Tevatron and the LHC with 8~TeV,  $\sigma_{t\bar t}^{(2)}$ and $\sigma_{t\bar t}^{(8)}$ respectively, including the
leading-order contributions of the top dipole moments.
Since the amplitudes contain at most two insertions of the anomalous vertices in
Eq.~(\ref{ec:lagr}), the dependence on $\da$ and $\dv$ can be parametrized by a fourth-order polynomial in these two variables.\footnote{We have made the choice of keeping all terms in the expansion of Eq.~\eqref{eq:fit}. Third-order and fourth-order terms do not play any role in the regions relevant for the combined limits. Quadratic terms correspond to contributions of order $\Lambda^{-4}$. Concerning the $\dv^2$ term, dimension-eight operators could generate additional contributions at the same $\Lambda^{-4}$ order, and including them would change the interpretation of the limits obtained on $\dv$. On the other hand, dropping quadratic and higher-order terms, or equivalently truncating the series at order $\Lambda^{-2}$, would imply to neglect effects that formally appear at order $\Lambda^{-4}$ but that are relevant in the extraction of the limits. We have therefore adopted the expansion of Eq.~(\ref{eq:fit}).} We find
\begin{eqnarray}
  \sigma_{t\bar t}^{(2)}(\text{pb})  & = &
  \sigma_\text{SM}^{(2)}(\text{pb}) - 45.5\,\dv + 131\,\dv^2 
  - 64.7\,\dv^3 \notag \\
& & + 55.5\,\dv^4  + 40.7\,\da^2  + 56.5\,\da^4 \notag \\
& &  - 66.2\,\dv\da^2 + 116\,\dv^2\da^2 \ , \notag \\
  \sigma_{t\bar t}^{(8)}(\text{nb}) & = & \sigma_\text{SM}^{(8)}(\text{nb}) - 1.53\,\dv + 10.1\,\dv^2 - 23.0\,\dv^3  \notag \\
& & + 28.6\,\dv^4  + 7.0\,\da^2 + 28.6\,\da^4 \notag \\
& & - 23.1\,\dv\da^2 + 57.3\,\dv^2\da^2 \ ,
\label{eq:fit}
\end{eqnarray}
when employing the NNPDF~2.3 set of parton densities~\cite{Ball:2012cx}. We
extract the limits on $\dv$ and $\da$ in Fig.~\ref{fig:tevlhc8} by using
the most recent total rate measurements at
the Tevatron, $\sigma_\text{exp}^{(2)}  = 7.60 \pm 0.41$ pb~\cite{Aaltonen:2013wca},
 and the LHC with 8~TeV, $\sigma_\text{exp}^{(8)} = 241.5 \pm 8.5$ pb~\cite{CMS:2014gta},
together with the most precise SM predictions at the
next-to-next-to-leading order accuracy in QCD,
$\sigma_\text{SM}^{(2)} = 7.35 \pm 0.21$ pb and
$\sigma_\text{SM}^{(8)} = 252.8 \pm 14.4$ pb~\cite{Czakon:2013goa}.
In the results displayed on Fig.~\ref{fig:tevlhc8}, we define the SM case where
there are no new physics contributions to the top dipole
moments by \mbox{$\dv=\da=0$}.
The shaded area corresponds to the overlap of the corresponding 95\% CL allowed
regions from the Tevatron (dashed) and the LHC (solid) measurements, and implies
that \mbox{$-0.012 \leq \dv \leq 0.023$} and \mbox{$|\da| \leq 0.087$}.
Our results are found compatible with the ones previously obtained in
Ref.~\cite{Hioki:2013hva} and given at the 68.3\% CL, and are
stronger than the bounds derived from the spin correlation
measurements~\cite{Bernreuther:2013aga,CMS:2014bea}.\footnote{After the completion of this work, an NLO calculation of the effects of a top chromo-magnetic moment has appeared~\cite{Franzosi:2015osa}. Despite the different setup used in the computation, the resulting limits on $\dv$ are rather similar to ours, and stable with respect to QCD corrections.}

With the important amount of $t \bar t$ data to be collected at the upcoming LHC
run with $pp$ collisions at 14~TeV, top dipole moments could be
probed by going beyond the use of inclusive total cross sections, and could
benefit from differential cross section measurements. For illustration,
we consider three representative cases and focus on inclusive cross section
measurements as well as on the production of $t \bar t$ pairs with an
invariant mass $\mtt$ required to be larger than either 1~TeV or 2~TeV. These
last two observables are expected to exhibit an enhanced
sensitivity to the top dipole moments because of the larger momentum
transfers that are now (phase-space) favored and the specific
Lorentz structure of the top dipole
operators in Eq.~\eqref{ec:lagr}. For the inclusive measurement, our predictions
rely on standard $t \bar t$ reconstruction techniques so that the $t \bar t$
signal is considered well separable from the background, in a way similar to
what has been achieved with collision data at 7~TeV
and 8~TeV. We
subsequently assume an overall uncertainty of 5\% on the would-be measurement
at 14~TeV. (For comparison, the 8 TeV measurement has a 3.5\% uncertainty.)

When the $t\bar t$
system has an invariant mass $\mtt > 1$~TeV or 2~TeV, the produced top quarks are
usually boosted. This renders any associated measurement more difficult
because of both the smaller statistics and the large QCD multijet
contribution where a pair of boosted top and antitop quarks is faked. In order
to realistically
estimate the uncertainty that would be associated with measurements in such
regimes, we restrict our analysis to $t \bar t$ pairs produced in the central
region of the detector (with a pseudorapidity satisfying $|\eta| < 2$). This
ensures a better performance of the top tagging algorithms due to a finer
detector granularity, so that one could aim for a better
rejection of the QCD background. We employ, in the $\mtt > 1$~TeV (2~TeV)
case, the third working point of CMS for top quarks with a transverse
momentum $p_T > 400$~GeV (800~GeV)~\cite{CMS:2014fya}, so that a boosted top
quark will be correctly tagged with an efficiency of about 12.5\%, for a
mistagging rate of a QCD jet as a top quark
of about 0.03\% (the studies of top-tagging performance by
ATLAS are documented in Ref.~\cite{TheATLAScollaboration:2013qia}). We present, in Table~\ref{tab:1}, the corresponding fiducial
cross sections for both the top-antitop signal and the
multijet background, together with the sensitivity defined as $\sqrt{S+B}/S$
where $S$ and $B$ are the numbers of signal and background events normalized to
a luminosity of 100 fb$^{-1}$, respectively.

\begin{table}
  \begin{center}
  \begin{tabular}{c|c|c||c}
    Invariant mass selection & $\sigma_{t \bar t}$ & $\sigma_{jj}$ & $\sqrt{S+B}/S$ \\
    \hline
    \hline
    $\mtt\; (\text{or } \mjj) > 1$ TeV & 1.0~pb & 0.89~pb & 0.004\\
    $\mtt\; (\text{or } \mjj) > 2$ TeV & 16~fb  & 40~fb  &  0.047\\
  \end{tabular}
  \caption{\small Fiducial cross sections for boosted $t \bar t$ and dijet
    production at the LHC (\mbox{$\sqrt{s}=14$~TeV}), after
    accounting for a centrality
    requirement and appropriate top tagging and misidentification rates. The
    sensitivities are normalized to 100 fb$^{-1}$ of simulated collisions.
    \label{tab:1}}
    \end{center}
\end{table}

\begin{figure}[t]
  \begin{center}
  \includegraphics[width=.80\columnwidth,clip=]{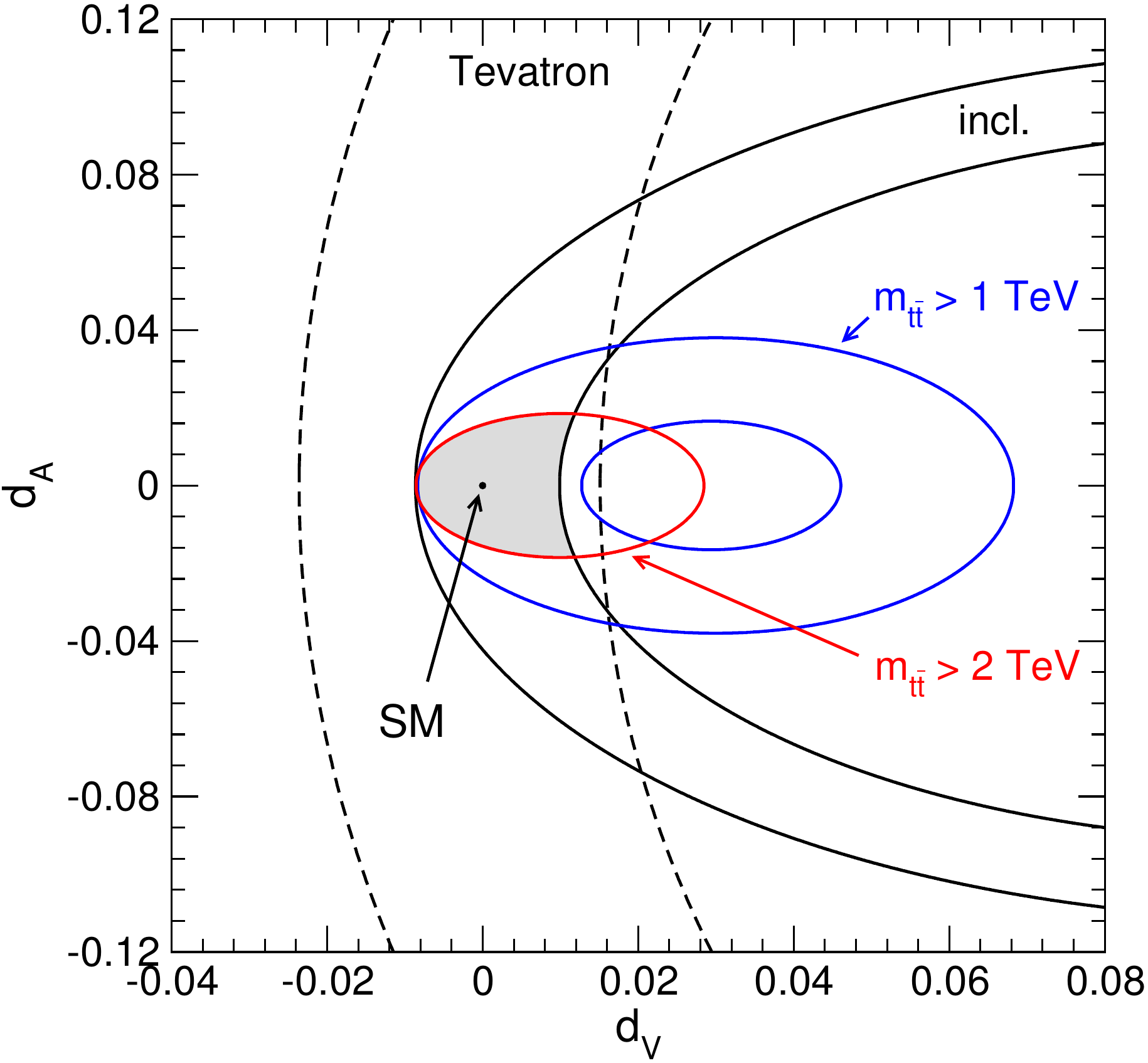}
  \caption{\small Expected 95\% CL limits on $\dv$ and $\da$ at the future LHC
    run, with \mbox{$\sqrt{s}=14$~TeV}. We show results using inclusive
    $t\bar t$
    cross sections (solid black) and after considering only top-antitop pairs
    with an invariant mass larger than 1~TeV (solid blue annulus) and
    2~TeV (solid red ellipse).
    For comparison, the Tevatron limit is also displayed (dashed).}
  \label{fig:lhc14}
  \end{center} 
\end{figure}

We use the above results to deduce the statistical uncertainties that would be
related to a fiducial cross section measurement in the large $\mtt$ region
for $pp$ collisions at 14~TeV. Adding in quadrature
systematic uncertainties, assumed to be 5\%,\footnote{While this figure may seem optimistic for the high-$\mtt$ tail measurement, one should also bear in mind that a simple counting experiment such as the cross section measurement above a given $\mtt$ cut may be improved by a differential measurement. In this case, the high-$\mtt$ tail is much more sensitive to top dipole moments than the region near threshold, which can be used for the reduction of the overall normalisation uncertainty.} we show,
in Fig.~\ref{fig:lhc14}, the limits expected to be extracted by using
inclusive cross sections (solid black) and after imposing that the top-antitop
systems under consideration have an invariant-mass larger than 1~TeV (solid
blue annulus) and 2~TeV (solid red ellipse). For comparison, we superimpose the limits
derived from inclusive cross section measurements at the Tevatron (dashed) while
the bounds derived from the first LHC run, already presented in Fig.~\ref{fig:tevlhc8}, are omitted as they are superseeded by
the potential LHC result at 14~TeV.
Aside from the fact that limits derived when $\mtt$ is required to be large
are more constraining, the shape of the allowed region in the
$(\dv,\da)$ plane changes for $\mtt > 2$~TeV, turning out to be
an ellipse instead of an annulus. This can be traced back to the smaller importance of the cubic terms $\dv^3$, $\dv \da^2$ in the expansion of the cross sections. For example, for $\mtt > 2$ TeV,
\begin{eqnarray}
  \sigma_{t\bar t}^{(14)}(\text{pb}) \!\!& \!=\! &\! \sigma_\text{SM}^{(14)}(\text{pb})
    \!-\! 1.43\,\dv + 75.1\,\dv^2 - 226\,\dv^3 \notag\\
 & &\!\! + 4410\,\dv^4 + 72.4\,\da^2 \!+\! 4410\,\da^4
      \!-\! 229\,\dv\da^2 \notag \\
      & &  \!\!+ 8830\,\dv^2\da^2 \,.
\label{eq:fit2}
\end{eqnarray}
Furthermore, despite the large coefficients, the quartic terms $d_V^4$, $d_A^4$ and $d_V^2 d_A^2$ are always subleading for
\mbox{$d_{V,A} \lesssim 0.03$}.

We estimate, with the shaded area in the figure, the 95\% CL bounds derived from
combining the inclusive and high-$\mtt$ measurements at the LHC run II. The top dipole moments are constrained to fulfill
\mbox{$-0.0086 \leq \dv \leq 0.012$} and \mbox{$|\da| \leq 0.019$}, so that the
future run of the LHC is expected to improve the limits on $\dv$ by a factor of
two. Moreover, the sensitivity to a $CP$-violating chromoelectric moment using a $CP$-even
observable --- such as the $t\bar t$ fiducial cross section at high $\mtt$ --- is found remarkably similar
to the expected one when measuring $CP$-odd triple product asymmetries
(\mbox{$|\da| \leq 0.02$}~\cite{Bernreuther:2013aga}). Finally, assuming the
Wilson coefficient $C_{uG\phi}^{33}$ to be of at most $4\pi$, one can translate
the bounds on $\dv$ and $\da$ into a lower limit on the new physics scale
$\Lambda$ that is found to be $\Lambda\gtrsim5$~TeV. (This ensures the
validity of the effective field theory approach used in Eq.~\eqref{ec:lagr}.) For smaller $C_{uG\phi}^{33}$ the limits on $\Lambda$ are correspondingly looser. 

\begin{figure}[t]
  \begin{center}
  \includegraphics[width=.7\columnwidth]{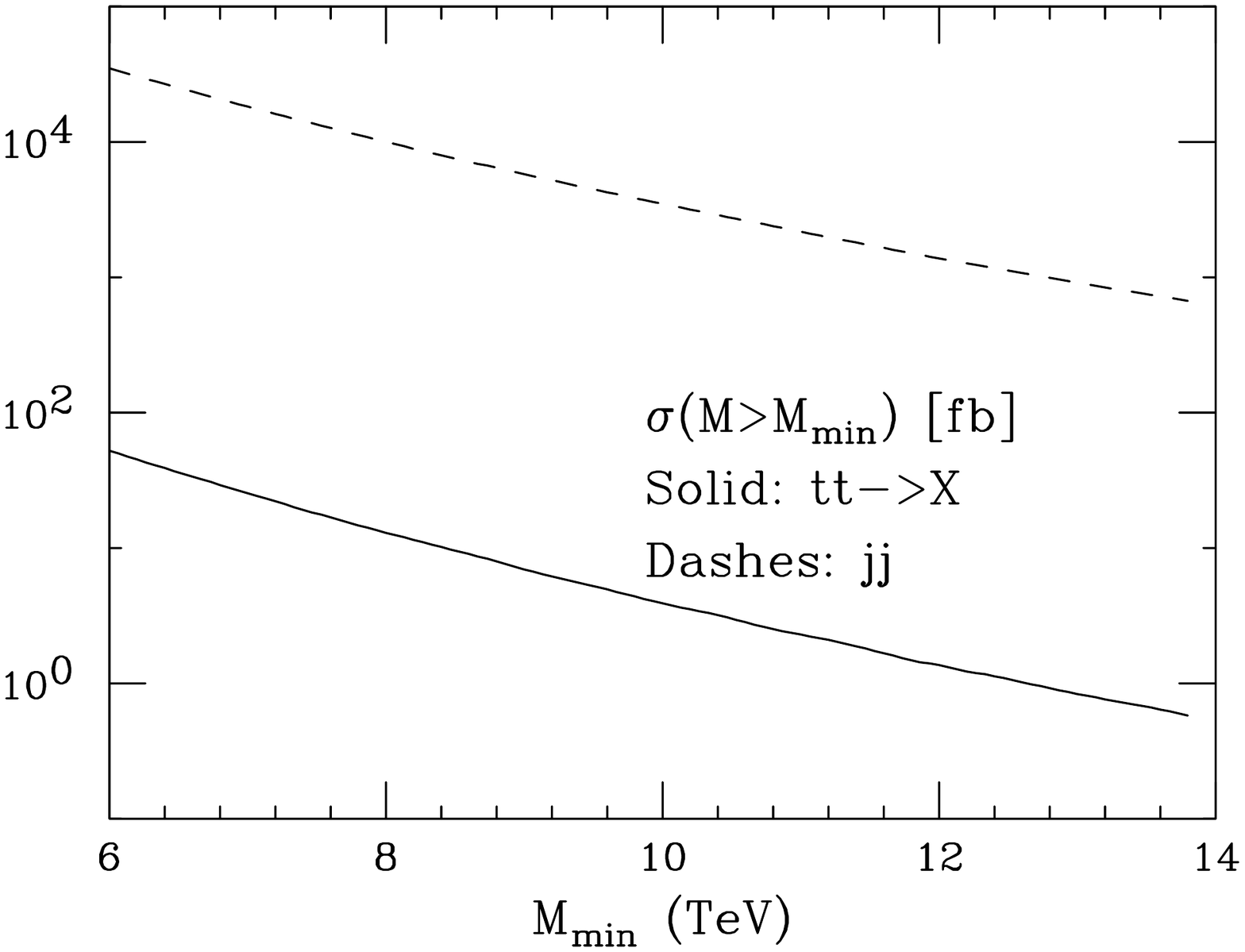}
  \includegraphics[width=.7\columnwidth]{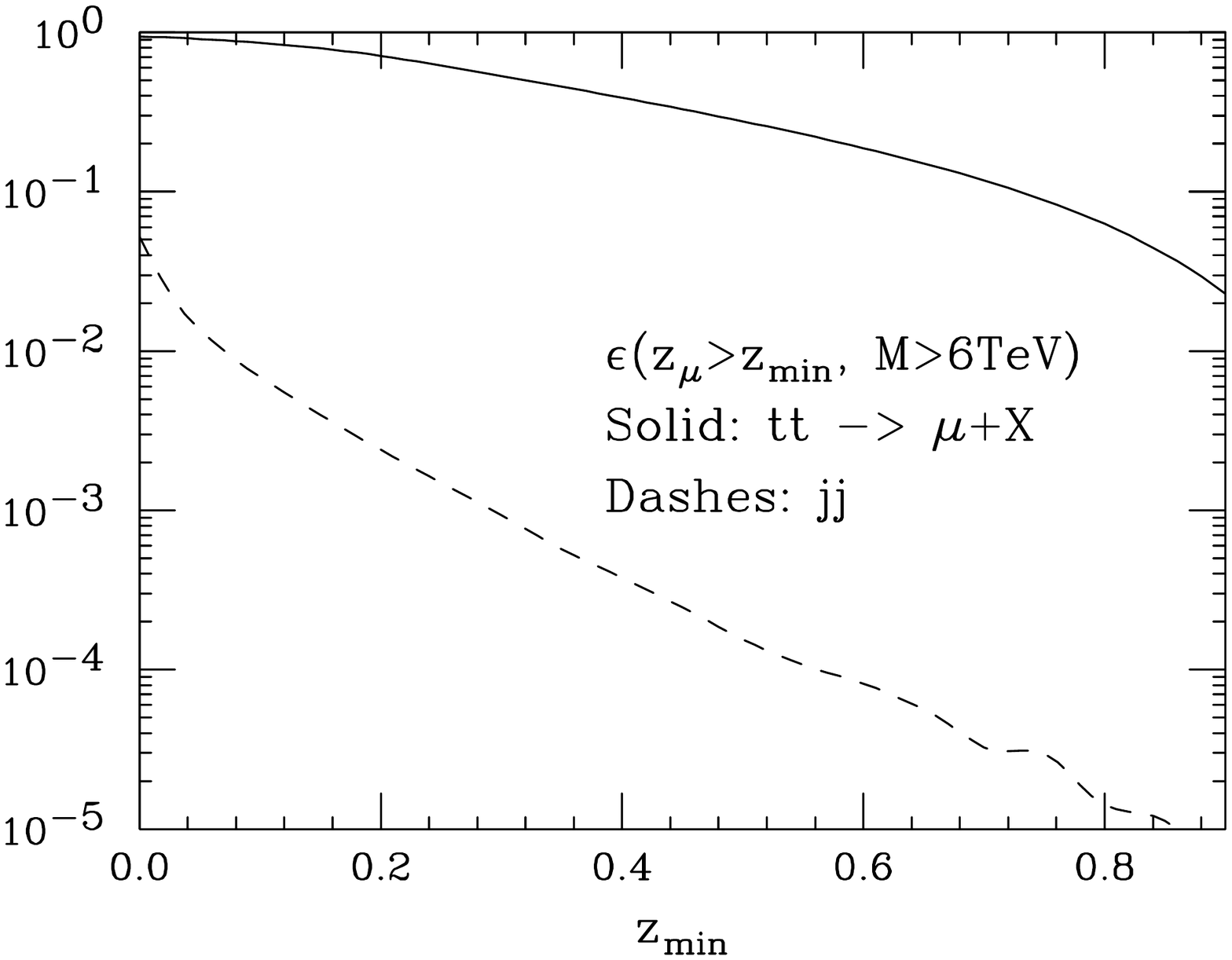}
  \includegraphics[width=.7\columnwidth]{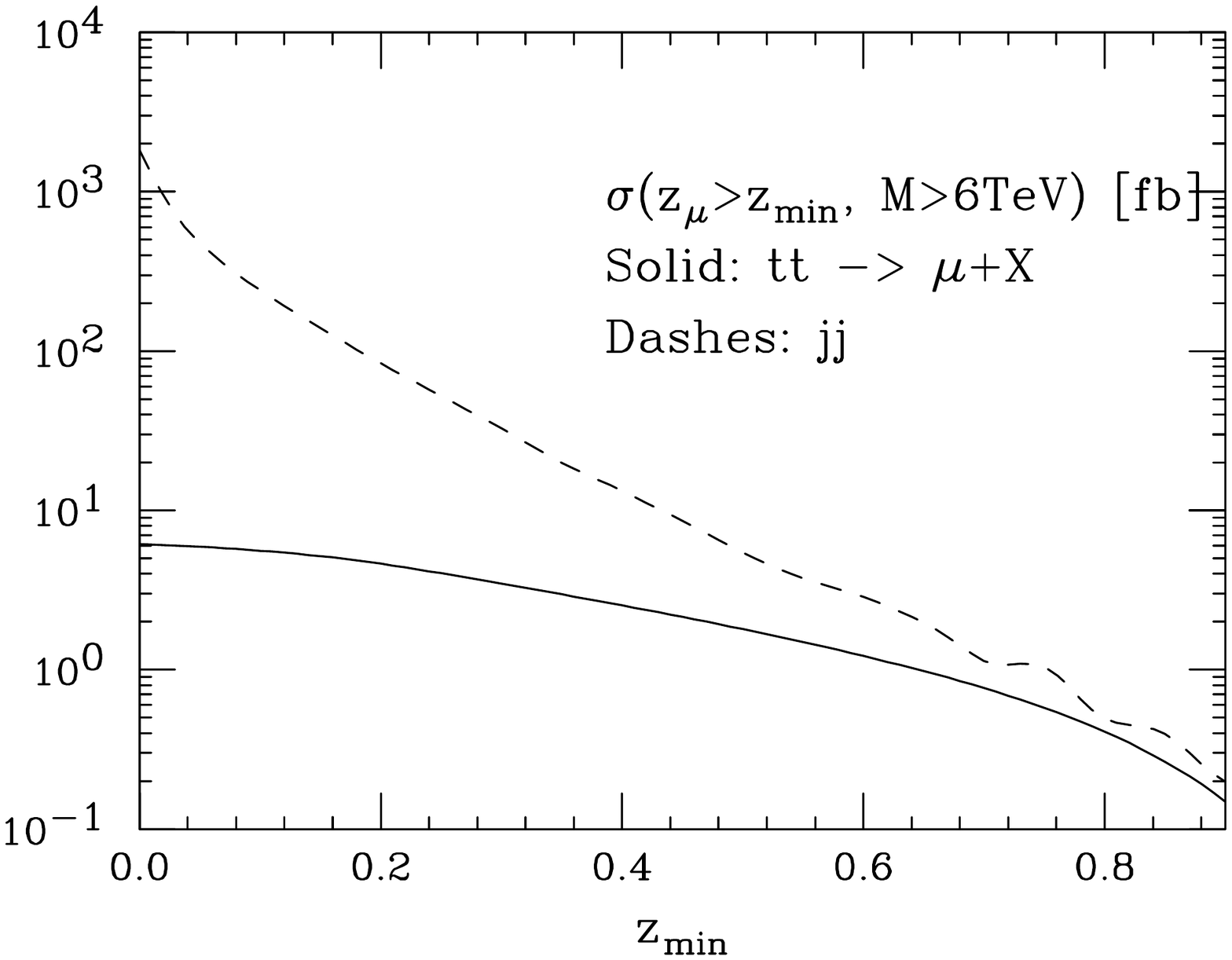}
  \caption{\small Top: cross sections for inclusive $t\bar{t}$ and
    dijet final states, with the mass of the two leading jets above
    $M_{min}$ (jets as defined in the text). Central: efficiency of the
    cut $z_\mu > z_{min}$, for $t\bar{t}\to \mu+X$ and for dijet
    events, with $M_{min}>6$~TeV. Bottom: $z_{min}$ dependence of the
    total rates for $t\bar{t}\to\mu+X$ and dijets, with $M_{min}>6$~TeV.}
\label{fig:top-jet}
  \end{center}
\end{figure}
\section{Sensitivity at 100~TeV}
A significantly large number of $t\bar t$ pairs with a multi-TeV invariant
mass are expected to be produced in several ab$^{-1}$ of $pp$ collisions at
100~TeV. With the opening of new kinematical regimes that have
never been probed so far, the performance of 
the standard boosted top
reconstruction techniques, developed in the LHC context and relying of
the top-jet substructure~
\cite{Englert:2014oea,CMS:2014fya,TheATLAScollaboration:2013qia,Thaler:2008ju,Kaplan:2008ie,Almeida:2008tp,Plehn:2009rk,Thaler:2011gf,Plehn:2011sj,Soper:2012pb,Schaetzel:2013vka,Backovic:2013bga,Larkoski:2014zma},
needs to be reassessed, also in view of the potential improvements in
the features of future detectors.
Considering that a top quark with $p_T=5$~TeV would have its three primary
decay products contained within a cone of $R \lesssim 0.05$, 
it is clear that the effectiveness of a tagging based on the jet substructure
would be strongly tied to the details of
the detectors, such as tracking performance and calorimetric
granularity\footnote{After completion of this
  work, a study appeared~\cite{Larkoski:2015yqa}, focused on the issue of
  tagging multi-TeV top jets. This paper confirms the potential for
  good tagging efficiency and background rejection, provided the
  detector performance can match the challenge of dealing with these
  high-density track environments.}. 
We therefore focus our study on a rather safe feature, in principle
usable with any conceivable detector design, which should
enable the differentiation of highly boosted top quarks from generic
QCD jets. That is the spectrum of muons coming from the top decays,
already discussed in the context of top tagging in
Refs.~\cite{Rehermann:2010vq,Brust:2014gia,Auerbach:2014xua}. We
show here that
the study of this observable provides a proof of principle that a
large-invariant-mass 
$t\bar t$ signal can be isolated from the otherwise overwhelming QCD
background, in spite of the limited efficiency due to the branching
ratio for a
muonic decay. It is likely that more efficient top taggers, along the
lines of what developed for the TeV regime of relevance to the LHC
will significantly improve the usable statistics and our final
sensitivity to anomalous top dipole moments.

We generate hard top-pair and dijet events with
\amc\ and further match them to the parton showering and hadronization
algorithms included in the {\sc Pythia}~8 program~\cite{Sjostrand:2007gs}.
We then consider all central ($|\eta|<2$) jets with $p_T > 1$~TeV
reconstructed by making use of an anti-$k_T$ algorithm with a
radius parameter $R=0.2$~\cite{Cacciari:2008gp}, as implemented in the
{\sc FastJet} package~\cite{Cacciari:2011ma}. We finally analyse the
reconstructed events with the {\sc MadAnalysis}~5 framework~\cite{Conte:2012fm}.
We preselect events featuring at least two reconstructed jets, and the
invariant mass 
of the system made of the two leading jets (generically denoted by $\mtt$)
is demanded to be greater than some threshold $M_{min}$.
In Fig.~\ref{fig:top-jet} (upper plot) we show the cross section as a
function of the minimum dijet invariant mass $M_{min}$, both for the
top signal and for the inclusive multijet background, which is over
two order of magnitude larger. 
As indicated above, in order to extract a top-antitop signal from the
multijet QCD background, we further require at
least one muon lying within a cone of $R=0.2$ around any of the
selected jets. This final step of the selection relies on the different
properties of the muons arising from multijet and $t\bar t$ events. In
the former case, they are found to only carry a small fraction of the
jet transverse momentum, as inferred by their production from
$B$-meson and $D$-meson decays, whereas in the latter case they are
induced by prompt top decays and can get a significant fraction of the
top transverse momentum. Events are consequently selected by
requesting a minimum value $z_\text{min}$ for the variable $z_\mu$, defined by
\be
  z_\mu = \max_{i=1,\ldots,n} \frac{p_T(\mu_i)}{p_T(j_i)} \,,
\label{eq:zz} \ee
where we maximize, over the $n$ muons possibly present in a given event,
the ratio of the muon transverse momentum $p_T(\mu_i)$ to the corresponding jet
transverse momentum $p_T(j_i)$. The efficiency of signal and
background, as a function of the requirement on $z_\mu$, are shown in the
central plot of Fig.~\ref{fig:top-jet}. In the case of the top signal,
we removed from the definition of this efficiency the trivial
branching ratio factor for the decay $t\bar{t} \to \mu+X$ (the
contribution from muonic decays of the $b$ hadrons is 
negligible in the relevant regions of $z_\mu$). 
Since the $z_\mu$ distribution has a slight dependence on
the transverse momentum of the jets, we show, as an example, the result averaged over the
set of events 
with dijet invariant mass larger than 6~TeV. Convoluting the
efficiencies with the appropriate rates, results in the cross sections
shown in the bottom plot of Fig.~\ref{fig:top-jet}. As we can see,
imposing $z_\mu \gtrsim 0.5$ reduces the
background by orders of magnitude, down to a level comparable to the
signal. Of course the experimental implementation of a selection like this
will require a good muon identification efficiency in the dense jet
environment, and a good
momentum resolution in the multi-TeV momentum range. The study of the
dijet rate, in the region of $z_{min}$ where the background dominates,
will provide nevertheless a good control sample for a robust
data-driven determination of the absolute background normalization.
\begin{table}
\begin{center}
\begin{tabular}{c| c| c | c}
    Invariant mass selection & $z$ selection & $S/B$ & ${\cal L}$\\
    \hline
    \hline
    $\mtt > $ 6~TeV  & $z_\mu>0.5$ & 0.39 & 36 fb$^{-1}$\\
    $\mtt > $ 10~TeV & $z_\mu>0.5$ & 0.74 & 200  fb$^{-1}$\\
    $\mtt > $ 15~TeV & $z_\mu>0.4$ & 0.25 & 2.4 ab$^{-1}$\\
\end{tabular}
\caption{\small \label{tab:fcc}Values of the selection threshold on
  the $z_\mu$-variable of Eq.~\eqref{eq:zz}
  for different invariant mass selections. We also present
  the related $S/B$ ratio and the luminosity ${\cal L}$ necessary
  for a $5\sigma$ extraction of a high-$\mtt$ signal
  from the multijet background.}
 \end{center}
\end{table}

To generate the following results, we considered the three reference
$M_{min}$ thresholds of 6, 10 and 15 TeV. The corresponding requirements on
the $z_\mu$-variable, which we used to optimize the signal
significance, are given in Table~\ref{tab:fcc}.  We also include here
the luminosities that are necessary for a signal extraction at the
$5\sigma$ level. We have verified that the results are similar when
using {\sc Herwig}++~\cite{Bahr:2008pv} instead of {\sc Pythia}, or
{\sc Alpgen}+{\sc Herwig}6~\cite{Mangano:2002ea,Corcella:2000bw}.

\begin{figure}[t]
\begin{center}
  \includegraphics[width=.80\columnwidth]{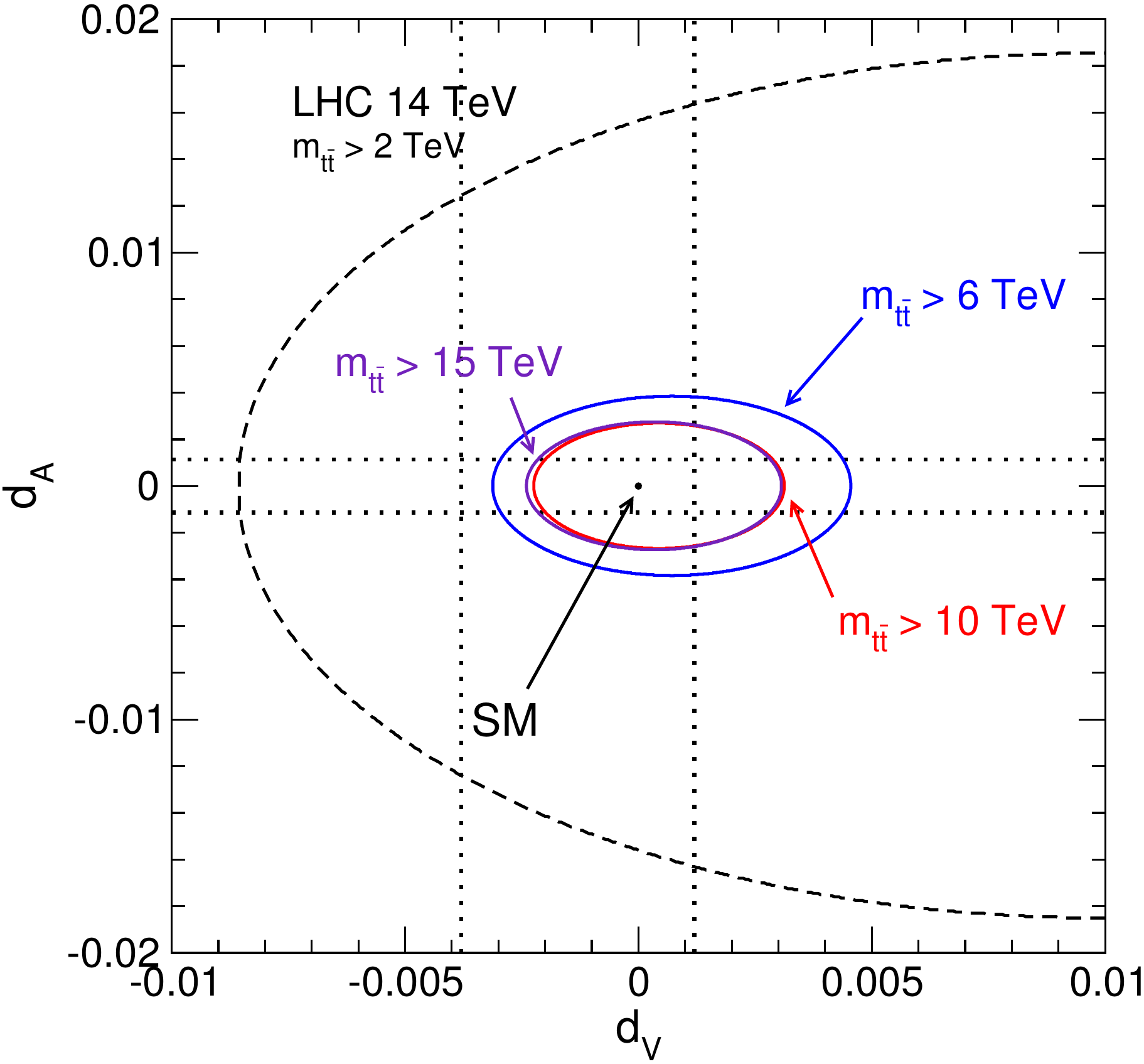}
  \caption{\small Expected 95\% CL limits on $\dv$ and $\da$ at
    100~TeV after
    considering top-antitop pairs with an invariant mass larger than 6~TeV
    (solid blue), 10~TeV (solid red) and 15~TeV (solid purple). For comparison,
    the expected limit from the second run of the LHC, with $\mtt>2$~TeV, (dashed line) and indirect limits (dotted lines) are also displayed.}
\label{fig:fcc}
\end{center}
\end{figure}

Using the above results to deduce the statistical uncertainties
associated with a would-be $t\bar t$ cross section measurement at
100~TeV, we additionally account for systematic uncertainties of 5\%
before deriving the expectation for constraining the top dipole
moments. (As previously discussed, we use a 5\% systematic uncertainty
as a reasonable reference value.) Assuming an integrated luminosity of
10~ab$^{-1}$, we display our results on Fig.~\ref{fig:fcc}, together
with the limits expected from the LHC run at 14~TeV when using $t\bar
t$ pairs with $\mtt>2$~TeV. As for the LHC case, the selection on the
top-antitop invariant mass enforces the contributions to the $t\bar t$
cross section that are quadratic in $\dv$ and $\da$ to dominate, so
that the allowed regions of the $(\dv,\da)$ plane are ellipses. As
also suggested by Table~\ref{tab:fcc}, the statistical uncertainties
start to be important for large $\mtt$ thresholds, so that the bounds
derived when $\mtt > 15$~TeV are similar to those obtained when $\mtt
> 10$ TeV. An optimal $\mtt$ selection would however strongly depend
on the boosted top identification efficiency and mistagging rate, and
may be different from the ones deduced in our simplified approach that
only aims to show that the observation of $t\bar t$ pairs at high
invariant mass can be envisaged. Enforcing $\mtt > 10$ TeV, the top
dipole moments are bound to $-0.0022 \leq \dv \leq 0.0031$ and $|\da|
\leq 0.0026$, which improves the LHC results by about one order of
magnitude, leading to constraints that are comparable to the indirect
ones obtained from $B$-decays and from the neutron electric dipole
moment. (In any case, direct and indirect limits are complementary,
since the latter are much more model-dependent and can be evaded with
additional new physics contributions.)  Conversely, these limits can
be translated in terms of a lower bound on the scale at which new
physics could be expected, which is found to satisfy $\Lambda \gtrsim
17$~TeV. This again ensures that the effective Lagrangian of
Eq.~\eqref{ec:lagr} is valid with respect to the magnitude of the
probed momentum transfers.

\section{Summary}
\begin{figure}[t!]
\begin{center}
\includegraphics[width=.80\columnwidth,clip=]{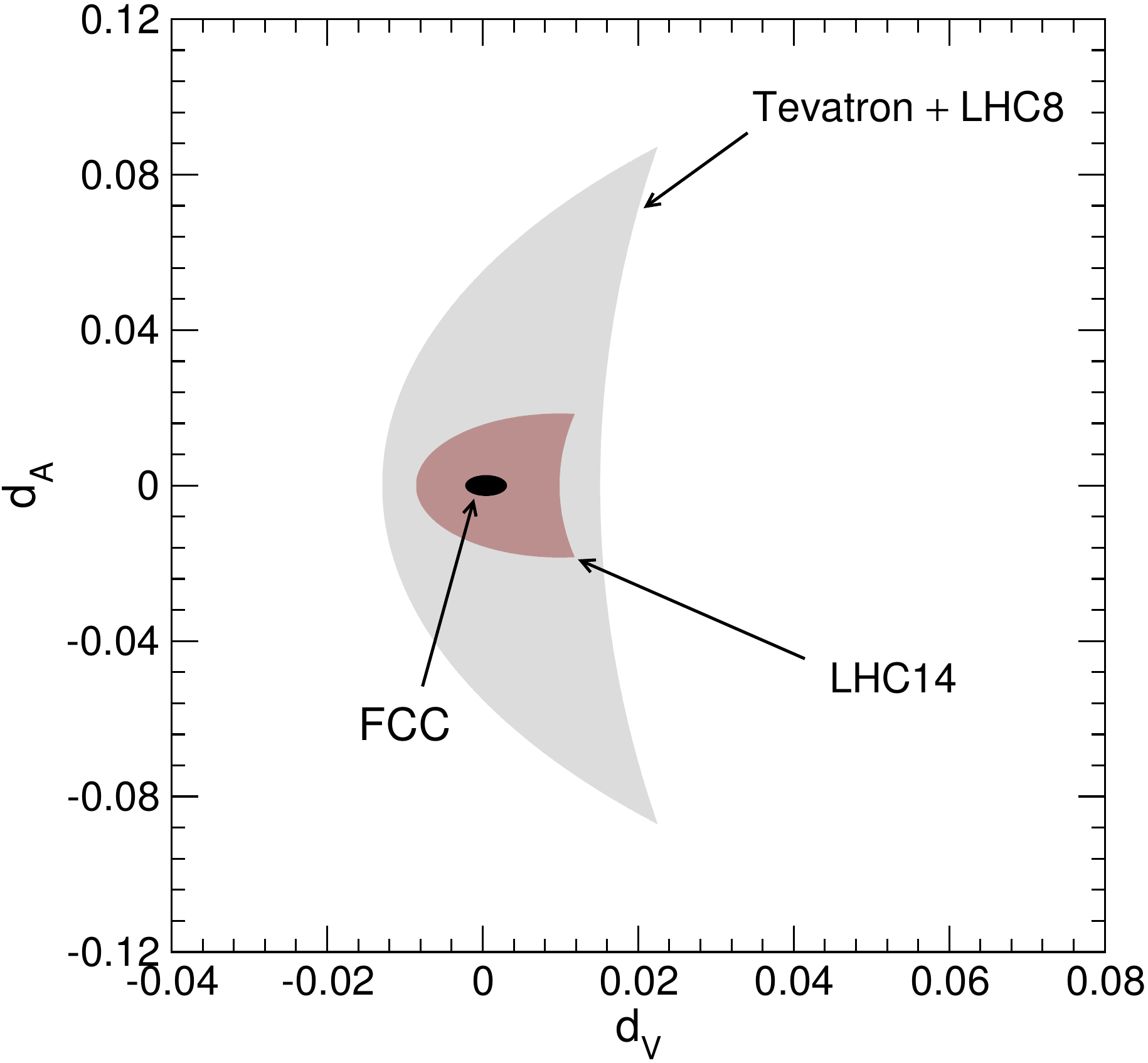}
  \caption{\small Comparison of current and expected limits on the top dipole
    moments at the LHC and at 100~TeV.}
\label{fig:comp}
\end{center}
\end{figure}

Direct limits on the top dipole couplings improve greatly by probing
higher momentum transfers, as a consequence of their Lorentz
structure. A 100~TeV $pp$ collider is therefore a suitable machine to
explore these anomalous interactions, in order to expose the indirect
effects of new heavy states. In this paper, we have investigated the
sensitivity of the future run of the LHC at 14~TeV and of a 100~TeV
collider to anomalous top chromoelectric and chromomagnetic dipole
moments $\dv$ and $\da$. We have considered both the study of $t\bar
t$ inclusive cross sections and of top-antitop pairs with a multi-TeV
invariant mass. The summary of our results is shown in
Fig.~\ref{fig:comp} where we compare the current direct bounds to the
projected limits for 100 fb$^{-1}$ at 14~TeV and 10 ab$^{-1}$ at 100
TeV.

The sensitivity at 100~TeV is expected to allow for a very important
improvement of the bounds derived from measurements at the Tevatron
and at the previous and future LHC runs, so that the $\dv$ and $\da$
allowed ranges could be reduced by more than one order of magnitude
with respect to the current direct limits. 
Furthermore, one may also wonder whether the expected LHC limits on
$\dv$ and $\da$, obtained from $t \bar t$ spin correlation
measurements and $CP$-odd triple product asymmetries, respectively,
could be improved at 100~TeV (at the LHC, they are of the same order
as the expected ones from cross section measurements in
Fig.~\ref{fig:lhc14}). Clearly, considering large energy scales --- by
selecting, for instance, high-$\mtt$ top-antitop pairs as done in this
work --- enhances the effect on the anomalous contributions, not only
in the fiducial cross section but also in the angular
distributions. However, the angular distributions that may be measured
in a typical LHC boosted top kinematical regime of $\mtt \lesssim 2$
TeV (see, \textit{e.g.}, Ref.~\cite{Baumgart:2011wk}) may not be
useful for ultra-boosted tops, and deserve further investigations.
Further studies would also be desirable to evaluate the
complementarity of the measurements discussed in this paper, with
those possible with $e^+e^-$ collisions at the top-antitop threshold,
where a large statistics is foreseen by the $e^+e^-$ option of the FCC
complex~\cite{Gomez-Ceballos:2013zzn}, and at higher energies
(\textit{e.g.}, at CLIC~\cite{Linssen:2012hp} or
ILC~\cite{Djouadi:2007ik}).

As a byproduct of our study, we developed and described a new robust
and effective approach to the problem of tagging top-antitop final
states of large invariant mass, exploting the hard muon spectrum in
top decays. We are confident that more detailed studies of top-tagging
algorithms, made possible also by the more aggressive detector
technologies envisaged for the future collider experiments, can
further improve our results.

What we showed is just an example of possible opportunities offered by
the huge samples of top quarks available at a 100 TeV $pp$
collider. Other areas that would certainly benefit include the study
of rare or forbidden decays ({\it e.g.} $t\to q+
g/Z/\gamma/H$ ($q=u,c$), tests of electroweak couplings (e.g. in
$s$-channel single top production $pp\to W^* \to t \bar{b}$ at very
large invariant mass), and high-precision measurements of production
asymmetries, spin correlations, etc. We look forward to future studies
addressing these observables. 
 
\begin{acknowledgements} This work was supported by the ERC grant
291377, ``LHCtheory: Theoretical predictions and analyses of LHC
physics: advancing the precision frontier'', by MICINN project
FPA2010-17915 and MINECO project FPA2013-47836-C3-2-P, by FCT project
EXPL/FIS-NUC/0460/2013, by the Junta de Andaluc\'{\i}a projects FQM
101 and FQM 6552, and by the Th\'eorie-LHC France initiative of the
CNRS/IN2P3.
\end{acknowledgements}

\end{document}